\documentclass[aps,twocolumn,showpacs,preprintnumbers,amsmath,amssymb,prl,reprint]{revtex4-1}

\usepackage[dvips]{graphicx}
\usepackage{dcolumn}
\usepackage{bm}
\usepackage{color}
\newcommand{\di}{\,\mathrm{d}}

\newcommand{\pdiff}[2]{\frac{\partial #1}{\partial #2}}
\newcommand{\Xv}{\mathbf{X}}
\newcommand{\Fv}{\mathbf{F}}
\newcommand{\Vv}{\mathbf{V}}
\newcommand{\Wf}[1]{W(\theta,#1,K)}
\newcommand{\Cf}[1]{C(\theta,#1,K)}
\newcommand{\gauss}{\mathcal{N}}

\begin{document}

\title{Multistable attractors in a network of phase oscillators with three-body interaction}

\author{Takuma Tanaka}
\email{tanaka.takuma@gmail.com}
\affiliation{
Department of Computational Intelligence and Systems Science,
Tokyo Institute of Technology, Japan.
}
\author{Toshio Aoyagi}
\affiliation{
Graduate School of Informatics,
Kyoto University, Japan
}
\affiliation{JST, CREST}

\date{\today}

\begin{abstract}
Three-body interactions have been found in physics, biology, and sociology.
To investigate their effect on dynamical systems, as a first step, we study numerically and theoretically a system of phase oscillators with three-body interaction.
As a result, an infinite number of multistable synchronized states appear above a critical coupling strength, while a stable incoherent state always exists for any coupling strength.
Owing to the infinite multistability, the degree of synchrony in asymptotic state can vary continuously within some range depending on the initial phase pattern.
\end{abstract}


\maketitle

\begin{figure}
\begin{center}
\includegraphics[width=8.6cm]{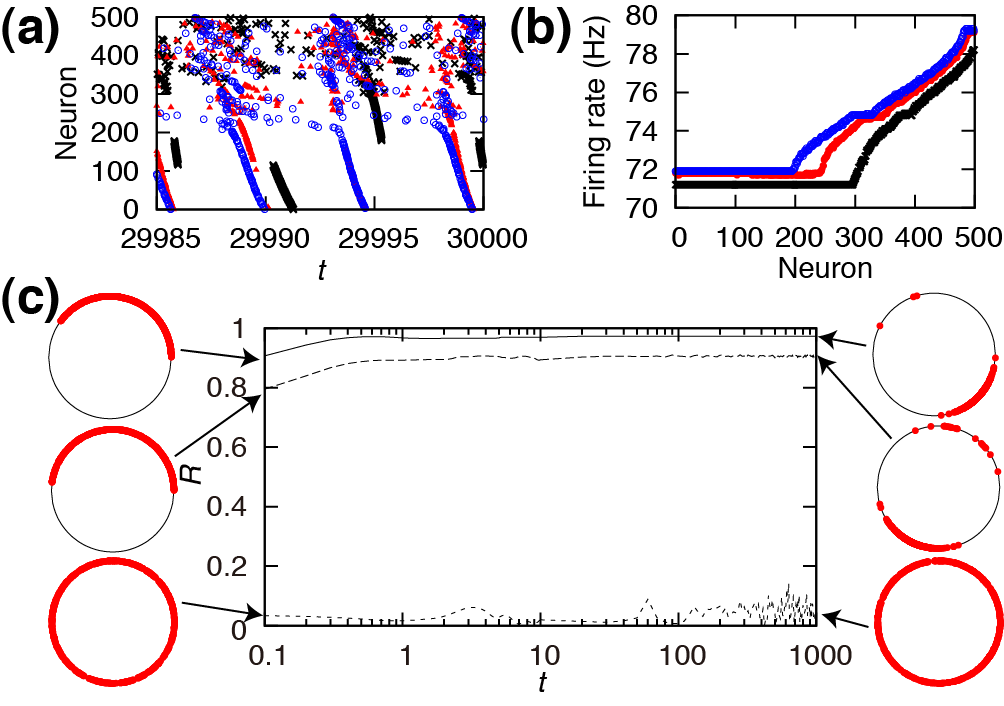}
\caption{\label{HHperturbed}
(color online).
Two examples of multistability arising from three-body interaction.
(a) Raster plot of the spikes in a network of the Hodgkin-Huxley neurons with short-term heterosynaptic plasticity ($N=500$).
Black, red, and blue dots represent the firings in the networks starting from different initial conditions.
Neurons are sorted in ascending order of the firing rate.
Firing rate of each neuron is shown in (b).
(c) Time evolution of the order parameter $R$ for three different initial conditions in the phase-oscillator systems with non-uniform random coupling ($N=500$).
Phase distributions of oscillators at $t=0$ and $t=1,000$ are shown on the left and right, respectively.
These two examples demonstrate that the same system can show different degrees of synchrony depending on the initial conditions.
}
\end{center}
\end{figure}

Interaction among particles or elements in classical mechanics, electromagnetism, and many other fields of physics is often modeled by two-body interaction.
Description by the linear superposition of two-body interactions has allowed us to predict the future orbits of the planets and to design drug molecules that tightly bind to the target protein.
However, it has been revealed that the net interaction experienced by an element cannot be written as the linear superposition of the two-body interaction in several systems, including physical systems \cite{Loiseau1967,*Buechler2007}, social and economic systems \cite{Shoham2009}, and neuronal networks \cite{Hsu1995,*Carter2007,*Larkum2009}.
A typical example is signal transmission from one neuron to another.
The signals are mediated by the release of neurotransmitters from synapses, and some neurotransmitters modulate the response of neurons to inputs from other neurons (heterosynaptic plasticity) \cite{Shepherd2004,*OConnor1994,*Saitow2005}.
This modulation can be regarded as three-body interaction,
although synaptic transmission is conventionally modeled as a two-body interaction.
To show what occurs in such neuronal networks with three-body interactions, we present a numerical simulation of a network of Hodgkin-Huxley neurons \cite{Dayan2001} with short-term heterosynaptic plasticity (see \footnote{
This model is described by
$\dot{V}_i = -g_\mathrm{Na}m_i^3h_i(V_i-E_\mathrm{Na})-g_\mathrm{K}n_i^4(V_i-E_\mathrm{K}) -g_\mathrm{Leak}(V_i-E_\mathrm{Leak})+I_\mathrm{syn}+I_i$,
$\dot{I}_\mathrm{syn} = -\frac{I_\mathrm{syn}}{\tau_1}+\frac{e}{N}\sum_{j,n}\delta(t,T_{j,n})\sum_k \cos\frac{2\pi[T'_i(t)-T'_k(t)]}{\tau_2}$,
where $I_i$ is the baseline input current of neuron $i$, $T_{i,n}$ is the $n$-th spike timing of neuron $i$, $T'_i(t)$ is the last spike time of neuron $i$ at time $t$, $\tau_1=3\mbox{~msec}$ is the decay time constant of synaptic current, $e=40\;\mu\mbox{A/mm}^2$ is the maximum amplitude of synaptic current, and $\tau_2=15\mbox{ msec}$ is the time scale of short-term plasticity.
The dynamics of gate variables follow those of the original Hodgkin-Huxley model \cite{Dayan2001}.
Baseline input follows $I_i=13+5(i/N-1/2)^3+5(i/N-1/2)$} for details).
In this model, the input from neuron $j$ to $i$ is modulated by the relative spike timing of neuron $i$ and other neurons in this model.
Figure~\ref{HHperturbed}(a) shows that this neuronal network exhibits multistability, in which
the numbers of synchronized neurons at the steady state vary depending on the initial conditions [Fig.~\ref{HHperturbed}(b)].
This seems to be a novel behavior not observed in systems with only two-body interactions.
However, this system is too complicated to show analytically why this multistability arises.

To analyze the neuronal networks with three-body interaction, we exploit the fact that neurons exhibit periodic firings in many cases.
Periodic activities are ubiquitous in not only neuronal networks, but also phenomena studied in other fields of biology, including
gene expression in \textit{E.~coli}, synchronous flashing of fireflies, and pedestrians' gait \cite{Danino2010,*Buck1988,*Strogatz2005}.
The behavior of these periodic activities is described by a form of phase oscillators in a quite general context \cite{Kuramoto1984,*Hoppensteadt1997,*Winfree2001,*Strogatz2000,*Acebron2005,*Ermentrout1996}.
However, three-body interaction among phase oscillators has not been studied yet.
Since phase oscillators are simple enough to be analytically tractable and structurally stable, theory of phase oscillators is a powerful tool in interpreting and elucidating complicated experimental results in which three-body interactions play an essential role.
In this Letter, we thus examine the effect of three-body interaction on the dynamics of globally-coupled phase oscillators.

As a natural extension of the system of limit-cycle oscillators with two-body interaction,
$N$-oscillator system with two- and three-body interaction is described by
$\dot{\Xv}_i = \Fv_i(\Xv_i)+\sum_{j,k} \Vv_{ijk}(\Xv_i,\Xv_j,\Xv_k)$,
where $\Fv_i$ describes the dynamics of uncoupled oscillator $i$ and $\Vv_{ijk}$ is the phase coupling function.
Two-body interaction $\Vv_{ij}(\Xv_i,\Xv_j)$ is then included as a special case of the three-body interaction $\Vv_{ijk}(\Xv_i,\Xv_j,\Xv_k)$.
Using the phase reduction technique, we can describe the dynamics of oscillator $i$ with one variable, phase $\phi_i$.
Thus, the dynamics of the system of phase oscillators with three-body interaction is generally given by
\begin{equation}
\dot{\phi_i} = \omega_i+\sum_{j,k} \Gamma_{ijk}(\phi_{ji},\phi_{ki}), \label{dynamics}
\end{equation}
where $\omega_i$ is the natural frequency of oscillator $i$, $\phi_{ji}=\phi_j-\phi_i$, and $\Gamma_{ijk}$ is the coupling function.

We present one example in which typical novel features arising from three-body interactions can be seen:
\begin{eqnarray*}
\dot{\phi_i}
&=&
\omega_i+\frac{1}{N}\sum_{j}
[a_{ij}\sin(\phi_{ji}+\alpha_{1ij})+b_{ij}\sin(2\phi_{ji}+\alpha_{2ij})]\\
&&+\frac{1}{N^2}\sum_{j,k}
c_{ijk}\sin(\phi_{ji}+\alpha_{3ijk})\cos(\phi_{ki}+\alpha_{4ijk}),
\end{eqnarray*}
where $a_{ij},b_{ij}\sim\gauss(0.3,0.01)$, $c_{ijk}\sim\gauss(6,4)$, and $\alpha_{1ij},\alpha_{2ij},\alpha_{3ijk},\alpha_{4ijk}\sim\gauss(0,0.09)$.
Here $\gauss(\mu,\sigma^2)$ denotes the normal distribution with mean $\mu$ and variance $\sigma^2$.
In all simulations throughout this Letter, the natural frequencies are drawn from $\gauss(0,1)$.
Some typical time evolutions of the order parameter $R$ representing the degree of synchrony is shown in Fig.~\ref{HHperturbed}(c), in which the above system starts from different initial conditions.
The order parameter $R$ is defined by
\begin{equation}
R\exp(\mathrm{i}\psi) = \frac{1}{N}\sum_j \exp(\mathrm{i}\phi_j), \label{orderparameter}
\end{equation}
where $\psi$ is the average phase associated with the order parameter.
As illustrated in Fig.~\ref{HHperturbed}(c), the system starting from a completely uniform initial distribution remains desynchronized, while the system with non-uniform initial distribution can go to various synchronized states in a similar way as in Fig.~\ref{HHperturbed}(a).
Two numerical simulations shown in Fig.~\ref{HHperturbed} suggest that the system containing three-body interaction can exhibit multistable behaviors in a structurally stable manner.

To investigate these behaviors analytically,
we here impose three assumptions which do not spoil the essence of the above dynamical behaviors.
First we assume that the phase coupling functions are identical for all oscillators, that is, $\Gamma_{ijk}(\phi_{ji},\phi_{ki})=\Gamma(\phi_{ji},\phi_{ki})/N^2$.
Second,  without loss of generality,
we can assume that the phase coupling function is symmetric, that is,
$\Gamma(\phi_{ji},\phi_{ki})=\Gamma(\phi_{ki},\phi_{ji})$,
because replacing the asymmetric coupling $\Gamma_\mathrm{asym}(x,y)$ with symmetric coupling $\Gamma_\mathrm{sym}(x,y)=[\Gamma_\mathrm{asym}(x,y)+\Gamma_\mathrm{asym}(y,x)]/2$ does not change the dynamics.
The last assumption is that inverting the phases of oscillators inverts the sign of forces among them, that is, $\Gamma(\phi_{ji},\phi_{ki})=-\Gamma(\phi_{ij},\phi_{ik})$.
Although this seems a rather tight constraint, this antisymmetricity is a property of the classical two-body coupling function $\Gamma(\phi_{ji})=\sin\phi_{ji}$.
We confirmed that the system under these constraints could exhibit the qualitatively same behavior as in Fig.~\ref{HHperturbed}(c).
Finally, we note that, owing to the $2\pi$-periodicity, $\Gamma$ can be approximated by the finite Fourier series
$\Gamma(\phi_{ji},\phi_{ki})
=
K_2(\sin\phi_{ji}+\sin\phi_{ki})/2+K_2'(\sin2\phi_{ji}+\sin2\phi_{ki})/2
+K_3(\sin\phi_{ji}\cos\phi_{ki}+\cos\phi_{ji}\sin\phi_{ki})$,
where $K_2$, $K_2'$ and $K_3$ are constants.
Thus, the dynamics of globally-coupled phase oscillators with this type of three-body coupling is given by
\begin{eqnarray}
\dot{\phi_i} &=& \omega_i+\frac{1}{N}\sum_{j}
(K_2\sin\phi_{ji}+K_2'\sin2\phi_{ji}) \nonumber\\
&&+\frac{2K_3}{N^2}\sum_{j,k}
\sin\phi_{ji}\cos\phi_{ki}. \label{fullform}
\end{eqnarray}

We further simplify this model equation to make it analytically tractable.
Using order parameter $R$ and setting $K_2=0$, $K_3'=0$, and $K_3=K$, we obtain the equations of dynamics with pure three-body interaction,
\begin{equation}
\dot{\theta_i} = \omega_i
-KR^2\sin 2\theta_i, \label{onebody}
\end{equation}
where $\theta_i=\phi_i-\psi$ is the relative phase of oscillator $i$ to the average phase $\psi$ [Eq.~(\ref{orderparameter})].  Because we are using a co-rotating frame, we may here assume that the average phase $\psi$ is constant.
We assume that the frequency of the average phase $\psi$ equals the mean of the distribution $g(\omega)$ of the natural frequency, the standard normal distribution.
This assumption simplifies the equations to be derived, and,
in addition, the solution of the derived self-consistent equation fits substantially well with the numerical results,
although this assumption may not hold in some cases.

\begin{figure}
\begin{center}
\includegraphics[width=8.6cm]{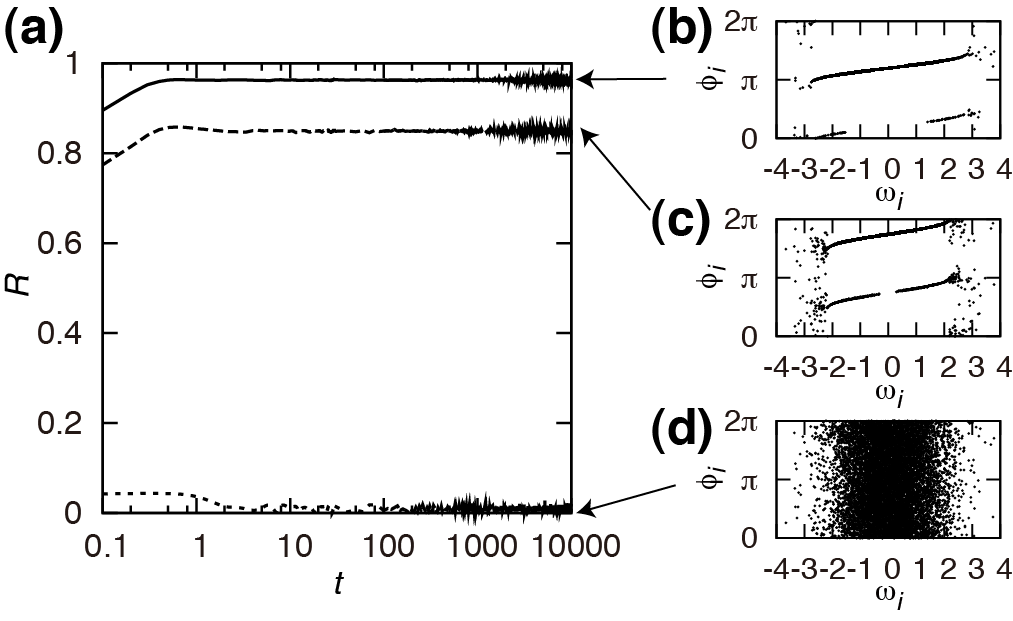}
\caption{\label{various}
(a) Time evolution of the order parameter $R$ from three different initial conditions in the mean field model with $N=10,000$ and $K=3$.
(b,c,d) $\omega_i$-$\phi_i$ relationship for different initial conditions at $t=10,000$.
}
\end{center}
\end{figure}

Numerical simulations of Eq.~(\ref{onebody}) with $N=10,000$ oscillators and $K=3$ from three initial conditions are shown in Fig.~\ref{various}(a).
Order parameter $R$ takes various values depending on the initial conditions.
Synchronized and desynchronized states coexist in the same parameter region.
The relationship between the natural frequency $\omega_i$ and the phase $\phi_i$ is also shown in Fig.~\ref{various}(b,c,d).
Figure~\ref{various}(c) indicates that oscillators can be phase locked to two specific phases.
Indeed, an oscillator with natural frequency $\omega_i$ can be phase locked to 
$\theta_i = \frac{1}{2}\arcsin\frac{\omega_i}{KR^2},\;\pi+\frac{1}{2}\arcsin\frac{\omega_i}{KR^2}$,
if
$-KR^2\le\omega_i\le KR^2$.
On the other hand, Fig.~\ref{various}(d) shows that the system with the same parameter values can exhibit a completely desynchronized state.

\begin{figure}
\begin{center}
\includegraphics[width=8.6cm]{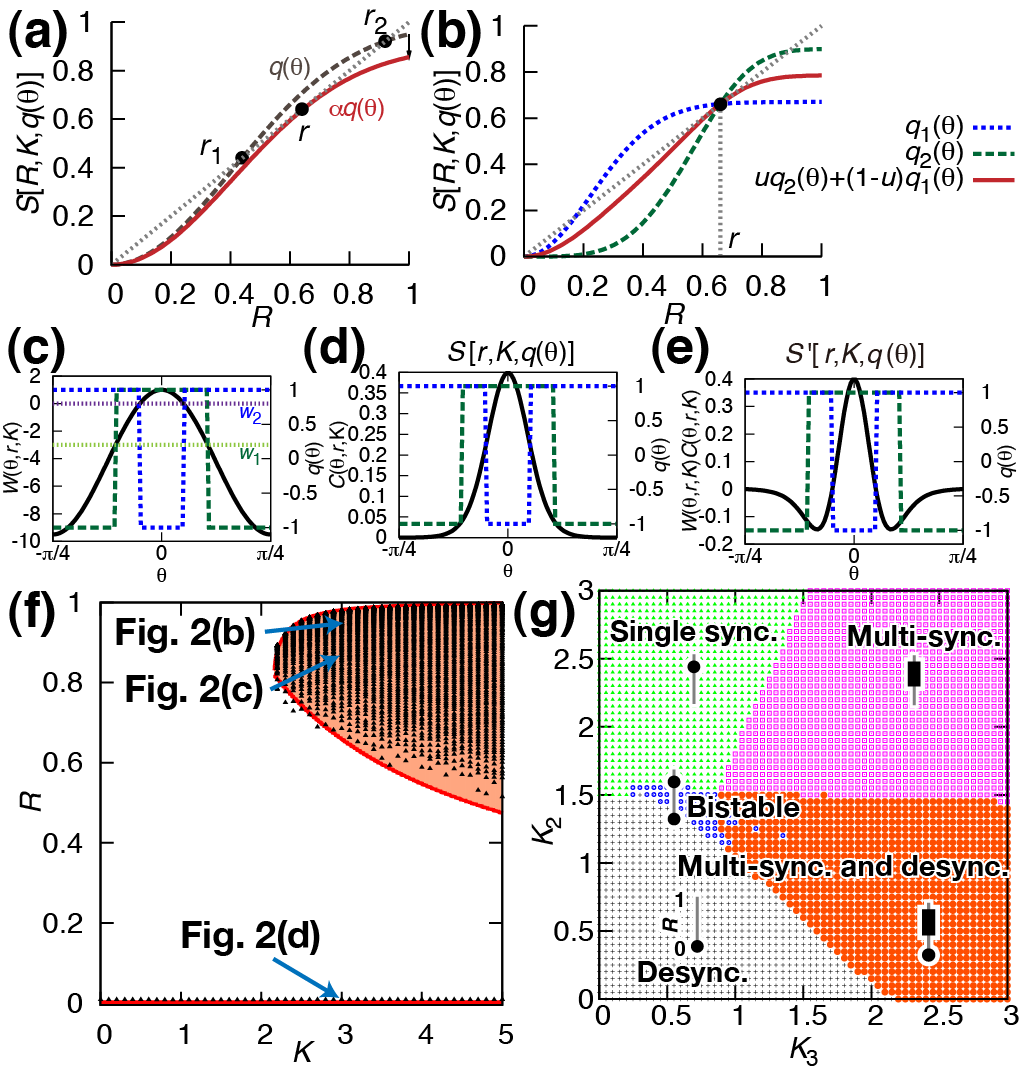}
\caption{\label{attainable}
(color online).
(a) For some $q(\theta)$, the self-consistent equation $R=S[R,K,q(\theta)]$ have two solutions, $r_1<r_2$ (brown line).
 Setting $\alpha$ appropriately makes $R=S[R,K,\alpha q(\theta)]$ have only one solution $R=r<r_2$ (red line).
Note that $S'[r,K,\alpha q(\theta)]=1$.
(b) If we have $S[r,K,q_1(\theta)]=r$ and $S'[r,K,q_1(\theta)]=s_1\le 1$ (blue line) and
$S[r,K,q_2(\theta)]=r$ and $S'[r,K,q_2(\theta)]=s_2\ge 1$ (green line), we can obtain $q(\theta)$ with which $S[r,K,q(\theta)]=r$ and $S'[r,K,q(\theta)]=1$ hold (red line).
(c,d,e) $q_2(\theta)$ which maximizes $S'[r,K,q(\theta)]$ (e) under the constraint $S[r,K,q(\theta)]=r$ (d) is given by $q_2(\theta) = 2\Theta[\Wf{r}-w_2]-1$ (green line) where $w_2$ is set to satisfy $S[r,K,q_2(\theta)]=r$ (c).
Under the same constraint, $S'[r,K,q(\theta)]$ is minimized by $q_1(\theta) = 2\Theta[w_1-\Wf{r}]-1$ (blue line) where $S[r,K,q_1(\theta)]=r$.
(f) Attainable region of the order parameter $R$ (orange region). 
Note that the incoherent state $R=0$ is stable for any $K$ (red line).
(g) Phase diagram of the system of phase oscillators when the strength of two-body and three-body interactions are changed.
The symbol in each region is a schematic representation of the attainable values of the order parameter $R$.
Gray lines represent the range of $R$ from 0 to 1.
The attainable values and ranges of $R$ are indicated by black circles and boxes, respectively.
}
\end{center}
\end{figure}

If all of the phase-locked oscillators are locked to $\theta_i = \frac{1}{2}\arcsin\frac{\omega_i}{KR^2}$, $R$ is given by
\begin{eqnarray}
R &=& \int_{-KR^2}^{KR^2}\cos\left(\frac{1}{2}\arcsin\frac{\omega}{KR^2}\right)g(\omega)\di\omega \nonumber\\
 &=& 2KR^2\int_{-\pi/4}^{\pi/4}\cos\theta\cos 2\theta g(KR^2\sin2\theta)\di\theta \nonumber\\
&\equiv& S(R,K), \label{selfall}
\end{eqnarray}
where we used
$\di\omega/\di\theta=2KR^2\cos 2\theta$,
and assumed that the non-phase-locked oscillators do not contribute to the value of the order parameter because the distribution $g(\omega)$ of natural frequency is the standard normal distribution.
The self-consistent equation $R=S(R,K)$ has a solution $R=0$ for any $K$.
In addition, $S'(0,K)=\pdiff{S}{R}|_{R=0}=0$ suggests that this solution is stable.
For some $K$, the self-consistent equation has a non-zero solution $R=r>0$ or two non-zero solutions $r_2>r_1>0$ [Fig.~\ref{attainable}(a)].
Equation~(\ref{selfall}) gives the order parameter of the system in which all of the phase-locked oscillators take $\theta_i = \frac{1}{2}\arcsin\frac{\omega_i}{KR^2}$.
Oscillator $i$, however, can also be phase locked to $\theta_i = \pi+\frac{1}{2}\arcsin\frac{\omega_i}{KR^2}$.
Defining $n(\theta)$ as the number of oscillators phase locked to $\theta$,
we characterize the distribution of the phase-locked oscillators with the function $q(\theta)=[n(\theta)-n(\theta+\pi)]/[n(\theta)+n(\theta+\pi)]$.
Note that $|q(\theta)|\le 1$.
Then, the order parameter of the system is given by
\begin{equation}
R = 2KR^2\int_{-\pi/4}^{\pi/4}q(\theta)\Cf{R}\di\theta \equiv S[R,K,q(\theta)], \label{selfpartial}
\end{equation}
where $\Cf{R}=\cos\theta\cos 2\theta g(KR^2\sin2\theta)$.

The largest attainable $R$ for coupling strength $K$ is given by the largest solution $r_2$ of Eq.~(\ref{selfall}), while the smallest attainable non-zero $R$  for the coupling strength $K$ is given by the minimum of the largest positive solution of the self-consistent equation Eq.~(\ref{selfpartial}) over all possible realizations of $q(\theta)$.
If $R=S[R,K,q(\theta)]$ has two non-zero solutions $r_2>r_1$, there exists $0<\alpha<1$ with which the largest non-zero solution of $R=S[R,K,\alpha q(\theta)]$ is smaller than $r_2$ because $S[R,K,\alpha q(\theta)]=\alpha S[R,K,q(\theta)]$ [Fig.~\ref{attainable}(a)].
Hence, to obtain the lowest attainable $R$, we have to find $q(\theta)$ with which $R=S[R,K,q(\theta)]$ has only one non-zero solution.
In other words, we find the smallest $r$ satisfying $S[r,K,q(\theta)]=r$ and $S'[r,K,q(\theta)]=1$ in varying $q(\theta)$, where
\begin{eqnarray*}
S'[r,K,q(\theta)] &\equiv& \pdiff{}{R}S[R,K,q(\theta)]\bigg|_{R=r}\\
&=& 4Kr\int_{-\pi/4}^{\pi/4} q(\theta) \Wf{r}\Cf{r}\di\theta,
\end{eqnarray*}
and
\[
\Wf{r} = 1+Kr^2\sin2\theta \frac{g'(Kr^2\sin2\theta)}{g(Kr^2\sin2\theta)}
\]
[Fig.~\ref{attainable}(c)].
To this end, first we fix $R$ to $r$ and examine whether there exists a solution $-1\le q(\theta)\le 1$ of the equations $S[r,K,q(\theta)]=r$ and $S'[r,K,q(\theta)]=1$.
If it exists, there is a solution $-1\le q_1(\theta)\le 1$ of equations $S[r,K,q(\theta)]= r$ and $S'[r,K,q(\theta)]=s_1\le 1$ [Fig.~\ref{attainable}(b), blue line], and there is a solution $-1\le q_2(\theta)\le 1$ of equations $S[r,K,q(\theta)]=r$ and $S'[r,K,q(\theta)]=s_2\ge 1$ [Fig.~\ref{attainable}(b), green line].
Conversely, if $q_1(\theta)$ and $q_2(\theta)$ are given,
$-1\le q(\theta)=u q_2(\theta)+(1-u)q_1(\theta)\le 1$,
where $0\le u=\frac{1-s_1}{s_2-s_1}\le 1$, is a solution of the equations $S[r,K,q(\theta)]=r$ and $S'[r,K,q(\theta)]=1$ [Fig.~\ref{attainable}{b}, red line].
Thus, the existence of $q_1(\theta)$ and $q_2(\theta)$ which satisfy $S[r,K,q_1(\theta)]=r$, $S'[r,K,q_1(\theta)]\le 1$, $S[r,K,q_2(\theta)]=r$, and $S'[r,K,q_2(\theta)]\ge 1$ is a necessary and sufficient condition of the existence of the solution of $S[r,K,q(\theta)]=r$ and $S'[r,K,q(\theta)]=1$.
It is sufficient for us to calculate the maximum and the minimum of $S'[r,K,q(\theta)]$ under the constraints $S[r,K,q(\theta)]=r$ and $|q(\theta)|\le 1$.

$S[r,K,q(\theta)]$ and $S'[r,K,q(\theta)]$ have the same domain of integration, and their integrands differ by a factor of $\Wf{r}$.
Hence, the maximum of $S'[r,K,q(\theta)]$ under the conditions $S[r,K,q(\theta)]=r$ and $|q(\theta)|\le 1$ is given by $S'[r,K,q_2(\theta)]$ where $q_2(\theta) = 2\Theta[\Wf{r}-w_2]-1$.
Here $\Theta(x)$ is the Heaviside function, and $w_2$ is set to satisfy $S[r,K,q_2(\theta)]=r$.
In this case, the phase-locked oscillators are distributed according to $n(\theta)/[n(\theta)+n(\theta+\pi)]=\Theta[\Wf{r}-w_2]$.
In other words, first we adjust $w_2$ to set $S[r,K,q_2(\theta)]=r$ [Fig.~\ref{attainable}(d)], and next we check whether $S'[r,K,q_2(\theta)]$ is larger than 1 [Fig.~\ref{attainable}(e)].
In the same way, we vary $w_1$ to set $S[r,K,q_1(\theta)]=r$, where $q_1(\theta) = 2\Theta[w_1-\Wf{r}]-1$, and check whether $S'[r,K,q_1(\theta)]$ is smaller than 1.

Thus, we have theoretically obtained the region of the order parameter which can be achieved by choosing suitable initial conditions [Fig.~\ref{attainable}(f), red line].
In this figure,  
the dots represent the data from numerical simulations ($N=10,000$).
The theoretical results agree with the numerical ones, though
several points with $K<3$ lie outside of the theoretically derived region.
This discrepancy may be because the system size is too small or because we assumed that the frequency of the average phase coincides with the mean of the distribution $g(\omega)$.

Finally, we should remark that interactions in real-world systems generally contain not only three-body but also two-body interactions.
We thus examine the behavior of the system described by
\[
\dot{\phi_i} = \omega_i+\frac{K_2}{N}\sum_{j}
\sin\phi_{ji}+\frac{2K_3}{N^2}\sum_{j,k}
\sin\phi_{ji}\cos\phi_{ki}.
\]
As Fig.~\ref{attainable}(g) shows, as $K_3$ increases, the system first starts out exhibiting either a single synchronized or desynchronized state depending on  $K_2$.  Then briefly, a small window in which these two states are bistable, appears.  Finally, multistable synchronized states, or for smaller $K_2$, a coexistence of desynchronized and multistable synchronized states [Fig.~\ref{attainable}(g), orange region] corresponding to the multistability shown in Fig.~\ref{various}, appears.
This implies that our theoretical result derived with pure three-body interaction is structurally stable and generic.

In this Letter, we have examined the behavior of phase oscillators with three-body interactions.
We have found that this system can take an infinite number of synchronized states in a structurally stable manner [Fig.~\ref{attainable}(g)].
We have derived the range of the order parameter $R$ that can be attained by varying the initial condition.
Our results are different from the chimera state \cite{Kuramoto2002,*Abrams2004}, because in our model we can continuously control the order parameter of the steady state by choosing the initial condition.
In addition, our model system can be completely incoherent even in the $K\rightarrow\infty$ limit (cf. \cite{Daido1996}).
There remain several questions to be answered.
Three-body interactions in real-world systems and their behavior should be compared to those of the present model.
Neurophysiological experiments \cite{Romo1999} have shown that some prefrontal neurons keep their level of activity for several seconds.
It is believed that this persistent activity serves as working memory by encoding an analog quantity in the firing rate of multistable neuronal networks.
Our results suggest the possibility that working memory uses the degree of synchrony among neurons to encode an analog quantity.
Finally, we should systematically investigate various types of coupling function and the dynamical behavior on complex networks \cite{Boccaletti2006}.

\begin{acknowledgments}
This work was supported by KAKENHI 21700250, 23115512, 19GS0208, 21120002, and 23115511 from MEXT, and Global COE Program ``Center for Frontier Medicine'', MEXT, Japan.
\end{acknowledgments}

\bibliographystyle{apsrev4-1}
\bibliography{letter}

\end{document}